\documentclass[10pt]{article}
\usepackage{graphics,epsfig}
\usepackage{harvard}
\usepackage{graphicx}
\usepackage{amssymb}
\usepackage{epstopdf}
\newcommand{\EE}{\hbox{\num{E}}}
\font\num=msbm10
\begin{document}
\title{A Case for Applying an Abstracted Quantum \\ Formalism to Cognition\footnote{Published as:
Aerts, D., Broekaert, J. and Gabora, L. (2003), A case for applying an abstracted quantum formalism
to cognition, in {\it Mind in Interaction}, ed. Campbell, R., John Benjamins, Amsterdam.}}
\author{Diederik Aerts, Jan Broekaert and Liane Gabora}
\date{}
\maketitle
\centerline{Center Leo Apostel (CLEA), Brussels Free University,}
\centerline{Krijgskundestraat 33, 1160 Brussels, Belgium.}
\centerline{diraerts@, jbroekae@ and lgabora@vub.ac.be.}
\bigskip
\bigskip
\bigskip
\noindent
%\section{Introduction}
This chapter outlines some of the highlights of efforts undertaken by 
our group to describe the role of contextuality in the 
conceptualization of conscious experience using generalized 
formalisms from quantum mechanics.
Conscious experience is filtered not just through innate categories
to give rise to stimulus-response reflexes,
but also learned categories, including concepts such as `container',
`democracy', `truth' and `falsehood'. The
meanings of these concepts are not rigid or static but shift fluidly
depending on context, increasing
dramatically our potential to both inform and be informed by the
world. As \citeasnoun[p. 101]{edelmantonini} put it: ``Every act of
perception is, to some extent, an act of creation, and every act of
memory is, to some degree, an act of imagination."

Elements of conscious experience, such as
perceived stimuli, retrieved memories, and
concepts are considered entities of the personal cognitive sphere,
referred to collectively as `conceptual
entities'.  These can be modelled by considering them as
configurations of properties, which are consistently testable
through personal and interpersonal cognitive processes such as
perception, reflection, and social interactions.
Specifically, we are concerned with the interface between the
conceptual entity and its extraneous surroundings: the
{\em context}. The interrelation and concatenation of concepts
consciously experienced as a stream of thought is
particularly affected by context, being influenced not only by the
ever-fluctuating associative structure of the
conceptual network but also by drives and emotion, as well as
environmental affordances including the social milieu.

One of the characteristic aspects of quantum mechanics is the {\em
measurement} context provoking an
indeterministic influence on the physical entity under consideration.
The mathematical formalism of quantum
mechanics describes precisely this influence and its corresponding
probabilities. The situation of contextual
influence in cognition is more complex than the one encountered in
the micro-world, but generalizations of the
mathematical formalisms of quantum mechanics are transferable to the
modeling of the creative, contextual manner
in which concepts are formed, evoked, and often merged together in
cognition \cite{gabora01,gaboraaerts01,gaboraaerts02}. Of
course, the nature of the conceptual entity differs essentially from
that of a physical entity in its transience and
subjectivity, and in its ontological status. The operational
formulation of `generalized' quantum mechanics, however, allows a
description of contextuality that does not specifically depend on the
physical nature of
the entities and contexts involved
\cite{aerts01,aerts02,aerts03,aerts04,aerts05,aertsaerts01,aertsaerts02,aertsetal01,aertsetal02}. 

For clarity, we emphasize that it is the {\em abstracted} formalism
which is `borrowed' from quantum theory, not in any way its
microphysical ontology of particles and fields. Our approach thus
concerns the formal structure of models that are able to describe
cognitive entities and processes with contextuality, not the
substrate that implements them in the brain.

\section{Formalization of Contextual Change}
In this section we explain why the mathematical modeling of a certain
kind of contextuality requires abstracted `quantum'
structures. Let us consider an entity---a conceptual
entity in our case---and a context--- e.g. some situation or social
framework. The entity, denoted
$S$, can be in different states. We represent the set of relevant
states of $S$ as the set $\Sigma$,
and denote individual states by symbols $p, q, r, \ldots \in \Sigma$.
The effect of the context, which we denote by
$e$, on the entity
$S$ is, in its most general form, that it changes the state $p$ of
the entity to another state $q$ of the entity. In
general this change of state under influence of the context is not
deterministic, which means that a probability of
change is involved. Let us denote by $\mu(q, p, e)$ the probability
that the context $e$ changes state $p$ of $S$
to state $q$. Of course, for a modelling of the
complete situation, not one but several different possible contexts must be
considered. Let is denote
the set of all relevant contexts ${\cal M}$, and specific contexts by
the symbols $e, f, g, \ldots \in {\cal M}$.
This means that the abstract form of the mathematical formalism we
consider is determined by the triplet
$(\Sigma, {\cal M},
\mu)$, where $\Sigma$ is the set of states of the entity $S$, ${\cal
M}$ is the set of contexts for $S$, and $\mu:
\Sigma \times \Sigma \times {\cal M} \rightarrow [0, 1]$, the
probability function that describes the transition
probability between different states of $S$ under the influence of a
specific context.

\subsection{Classical and Quantum Contextuality}
\label{sec:classicalquantumcontext}
The abstracted formalism gives us qualitative indications and
quantitative measures of contextuality, tools with which we can
evaluate and describe contextuality in various cognitive processes.
As in quantum mechanics, there are two kinds of contextuality in
cognition. In order to elaborate them, however, it is necessary to
present a few more details---albeit a strict minimum---of the
generalized quantum formalism.

Within the standard quantum mechanical formalism, two kinds of change
under the influence of context are distinguished. First there is a
{\em nondeterministic} change of state under influence of the
measurement context. For this change, each state
$p_u$ of the quantum entity $S$ under consideration is described by a
unit vector $u$ of a complex {\em Hilbert space} ${\cal H}$. Let
us denote this set of states by $\Sigma_{\cal H}$. Each (measurement)
context $e_H$ is described by a self-adjoint operator $H$ on
this Hilbert space ${\cal H}$. Let us denote the set of contexts by
${\cal M}_{\cal H}$. The probability function $\mu$ is given
as follows. Suppose the quantum entity is in state $p_u$ under the
influence of context $e_H$. The state $p_u$ is then
changed to one of the eigenstates $p_v$ of the self-adjoint operator
$H$, and the probability of change is given by:
\begin{equation}
\mu(p_v, p_u, e_H) = |\langle u, v \rangle|^2
\end{equation}
where $\langle u, v \rangle$ is the scalar product of vectors $u$ and
$v$ in the Hilbert space ${\cal H}$.

Second there is a {\em deterministic} change, which is the `mechanic'
evolution. It is described by the
Schr\"odinger equation, governing continuous change over time. For
clarity, we mention that, within a classical mechanical
formalism, only deterministic change under the influence of context
is possible, and it is described by the equations of Newton.

To a first approximation we can say that the term classical
contextuality refers only to situations where context works in a
deterministic way on the state of the entity under consideration,
while quantum contextuality includes nondeterministic change for
measurement contexts and deterministic change for dynamical contexts.

\subsection{The Statistical Situation}

Of course in classical physics, models are built
that include indeterminism, namely statistical mechanical models.
This indeterminism, however, is of a specific and limited
nature: it describes a {\em lack of knowledge} about the exact (pure)
state of the physical entity under
consideration. Thus, the notion of {\em statistical state} (or mixed
state) is introduced. In our formalism it is similarly possible to
describe entities (even conceptual entities) in statistical states.
In this statistical case, the distinction between classical and
quantum is subtle and depends on the structure of the probability
model involved. If the probability model is Kolmogorovian---meaning
that it satisfies Kolmogorov's axioms, and hence the event space is a
$\sigma$-algebra---only the classical statistical situation can be
described.

It is known that the quantum probability model is not Kolmogorovian
\cite{accardi01,accardifedullo01,pitowsky01}. It is possible to prove
that the nonKolmogorovian nature of the quantum probability model is
due to a lack of knowledge concerning how context interacts with the
entity under consideration, {\it i.e.} by the presence of {\em
fluctuations} in the interaction between context and entity. Even if
we were to suppose that at the ontological level the interaction
between context and entity engenders a change of state that is
deterministic, a lack of knowledge about this interaction gives rise
to a probability model that does not satisfy Kolmogorov's axioms.
Hence a quantum-like probability model is needed to model this
situation \cite{aerts03,aerts06,aerts08}.

Because of the confusion often created by this subtility, we will
elaborate it a little further. If a model contains only one state of
an entity and one context influencing this state, as is often the
case in an isolated
problem where no attempt is made to deliver a full model description
of the entity, the distinction between
classical and quantum described above vanishes. Consider the
situation of `casting a die'. In probability theory, only one state
and one context are given. Indeed, suppose we do not know exactly how
the die is cast---as is the case in the standard probability model of
this situation---then
this lack of knowledge gives rise to one state, a statistical state
that describes the situation ``the die is cast", without any
further specification of how it is cast. There is just one context,
for example the physical state of the table on which the
die is cast.  But we can inverse the description, and for example
consider the state ``the die is located within a cup that stands on
the table", which is a `pure' state as compared to the former
statistical state that we considered. Then the context becomes more
intricate: ``we, who take the cup, move it around, and cast the die
on the table, plus the situation of the table". Here the lack of
knowledge concerns the context, not the state of the die. Obviously
both situations give rise to the same
mathematical model, and Kolmogorovian probability theory can be applied.

The die example shows that for a very simple situation that can be
modeled using only one state and
one context, the mathematical structure of the probability model
cannot give us the information needed to distinguish between the two
ontologically different situations: lack of knowledge about the state,
and lack of knowledge about the context. Moreover, the
situation can always be captured within a Kolmogorovian probability
model. We note this explicitly because
in cognition as well as in other fields where probability theory is
used, the more complex situation is often cut down to a sample of
fragmented simple situations, where only one state and one context
are considered. Then each simple situation can be modeled by
standard probability theory within a Kolmogorovian structure. It is
only when the more complex situation of an entity changing and
evolving over a range of different states is considered that the
ontological difference between lack of knowledge of the state
giving rise to a Kolmogorovian structure, and lack of knowledge of
the context giving rise to a quantum structure, is revealed.
This phenomenon is well known in axiomatics. Indeed, the
Kolmogorovian structure of a probability model is an axiomatic
structure, and often ontological differences are only revealed by
means of the axiomatic structure if the situations considered are
not too simple (not too small). Often in cognition and other fields
of science, a complex situation
is cut into fragmented simple situations without paying attention
afterwards to how these fragmented situations are
pasted back again into a complex model that still corresponds to
reality. We believe that this at the origin of the fact that the
quantum nature---or
at least nonKolmogorovian nature---of such complex situations was not
identified much earlier. It has been proven that
in the case of an entity with simple dual contexts (contexts that
allow only two possible effects of change), at least three contexts
are needed for the resulting quantum structure to be revealed
mathematically \cite{accardi01,accardifedullo01}.

\section{Contextuality in Cognition}
In cognition, we will always lack
knowledge concerning the interaction between the context and the conceptual
entity. Hence the situation of the
presence of uncontrollable fluctuations on this interaction is the
standard one that we have to consider. This is the strongest
argument we put forward for the necessity of applying the mathematical
structures developed for generalized quantum mechanics to the
cognition. Again we remark that the quantum
structure of the mathematical model will only be revealed if more
than one state and one context are considered. A detailed analysis of
this situation, taken into account the effect of the presence of
fluctuations on the interaction between context and entity, giving
rise to a generalized quantum structure and nonKolmogorovian
probability model, can be found in \citeasnoun{gabora01}, and 
\citeasnoun{gaboraaerts02}. Prior to this general approach, specific 
abstracted quantum
models had been worked out for various cognitive situations by
our group. Indeed, it was already
obvious to us that quantum structures had to account for
the phenomenon of contextualization.

At present, we are far from a generally applicable quantum-like
approach to describing what happens in the mind. However, the
effectiveness of applying appropriate adaptations of the quantum
formalism to specific problems in the field is considerable. In the
next few sections, we illustrate how some of perplexing problems in
the formal description of human conscious experience can be
handled in this way. In each case, we explain how non-classical
features arise with respect to the relevant cognitive entities and
processes and how their contextual aspects are described through the
introduction of quantum-like formalisms.

\subsection{Violation of Bell Inequalities by Entangled Concepts}

The presence of entanglement---{\it i.e.}genuine
quantum structure---can be tested
for by determining whether correlation experiments on a {\em joint}
entity violate Bell inequalities \cite{bell01}. \citeasnoun{pitowsky01} proved
that if Bell inequalities are satisfied in an experiment, it follows
a classical Kolmogorovian probability scheme. The probability can
then be explained as being due to a lack of knowledge about the
precise state of the system in a classical manner. If, however, Bell
inequalities are violated, Pitowsky proved that no such classical
Kolmogorovian probability model exists. In that case the interwoven
nature of the compound entity is exposed.

Bell inequalities have been successfully applied to an elementary cognitive
setting \cite{aertsetal03}, with the inequalities functioning as a 
quantitative indicator of {\em
entanglement}. The application in cognition requires a brief 
presentation of basic
elements of the formalism.
We need to introduce four experiments $e_{i}$ each having only two 
possible outcomes, with $i$ running from
1 to 4. Furthermore, we need the expectation values $\EE_{ij}$, with $ i,
j $ each running from 1 to 4, for  coincidence experiments, {\em
i.e.}joint experiments of the type $e_i e_j$. The expectation values
are defined as:
\begin{equation}
\EE_{ij} =   P(o_i(u), o_j(u))+P(o_i(d), o_j(d))  - P(o_i(u),
o_j(d))- P(o_i(d), o_j(u))
\end{equation}
where  originally $u$ stands for `up' and $d$ for
`down', but actually they can stand for any outcomes out of the two 
possibilities.
So `yes' and `no' are valid as well. We assume that the associated
outcome values $o$ are either +1 (for $o(u)$) or -1 (for $o(d)$) ,
the correlation outcome function $P$  are then +1 or -1 as well.
Finally, from the assumption that the correlation $\EE_{ij}$ is
{\em local}, Bell derived his inequalities:
\begin{equation} \label{bellineq}
|\EE_{13} -\EE_{14}|+|\EE_{23}+
\EE_{24}|  \leq 2
\end{equation}
The essential point here is that when this inequality holds it 
indicates that the entity is not entangled; that is, the aspects of
the entity exposed by the different experiments are not
interdependent.

We illustrate its cognitive application as follows. Let there be a
person---the respondent in our experiment---who is very acquainted
with two cats Glimmer and Inkling which sometimes sport tinkling
bells on their necklaces.  Suppose now that we submit our respondent
to some basic experiments, which  are nothing more than asking for a
reply to one of the questions,\\
\indent {\em D} :  ``Think of one of
the cats, which one?" ({\em Directive} question)\\
\indent {\em S} :
``Do you hear its bell ring?" ({\em Sensory} question)\\
while at the
same time {\em  one of the cats---which first were  out of
sight---can jump on the scene.}
In the coincidence experiments we
consider combinations of the basic ones just mentioned, with of
course the possibility here that both cats appear in front of the
respondent.

\noindent We can summarize the basic experiments and
outcomes  as follows:\\

\medskip
\begin{center}
\begin{tabular}{l|l|cc}
              &    Experiments           & Outcome {\em up} &   Outcome
{\em down} \\
\hline
$e_1$ &  {\em D} + ``Glimmer appears"  &   ``Glimmer!"  & ``Inkling!" \\
$e_2$  &  {\em S} + ``Inkling appears"   &   ``ring"  &  ``silence"
\\
$e_3$  & {\em D} + ``Inkling appears"   &   ``Inkling!"   &
``Glimmer!" \\
$e_4$  & {\em S} + ``Glimmer appears" &  ``ring"  &
``silence"  \\
\end{tabular}
\end{center}
\medskip

\noindent The necessary conditions such
that Bell inequalities are violated in this experiment are as
follows.
As a  fact we must have {\em i)} both cats are wearing bells
around their necks that day.
We require the respondent {\em ii)} to
abide the question experiments, {\em iii)} to have  an activated
categorical concept `cat'. Most crucially is however the feature of
yielding to the coercive nature of internalized context {\em iv)}
when Glimmer is seen, there is a change of mind state and the
instance `Glimmer' is reported, and when Inkling is seen `Inkling' is
reported.
Then the outcome table for joint experiments in this
configuration can be immediately obtained:

\medskip

\begin{center}
\begin{tabular}{l|l|cccc}
      &    Experiments  & $o_i(u) o_j(d)$ & $o_i(d) o_j(u)$ & $o_i(u)
o_j(u)$ &  $o_i(d) o_j(d)$\\
\hline
$e_{13}$  & {\em D} + ``Glimmer\&Inkling" &   ``Glimmer!"  & 
``Inkling!"  & - & -\\
$e_{14}$  & {\em D} + {\em S} + ``Glimmer"   &  -  &  -  & ``Glimmer!+ring"
& - \\
$e_{23}$  & {\em D} + {\em S} + ``Inkling"   &   -  & -  &
``Inkling!+ring"& - \\
$e_{24}$  & {\em S} + ``Glimmer\&Inkling"  &   - &  -  &
``rings"  & -\\
\end{tabular}
\end{center}
\medskip
\noindent
This leads the to the correlation values:
$\EE_{13} = -1$. $\EE_{14} = +1$, $\EE_{23} = +1$, $\EE_{24} = +1$.
As a consequence we have the violation of the Bell inequality:
\begin{equation}
| \EE_{13} - \EE_{14} | + | \EE_{23} + \EE_{24} | = +4
\end{equation}
The reason that Bell inequalities are violated is that the respondent's state
of mind changes from activation of the abstract
categorical concept `cat',  to activation of either `Glimmer' or
`Inkling'. We can thus view the state `cat' as an entangled state of
these two instances of it. The relationship between a concept and
specific instances of it with a coercive context lead to the 
violation of the Bell
inequality \cite{aertsetal03}. Thus we have evidence that this 
formalism reflects the
underlying structure of concepts. In \citeasnoun{aertsetal03} we
show that this result is obtained because of the presence of EPR-type
correlations amongst the features or properties of concepts. The EPR
nature of these correlations arises because of how concepts exist in
states of potentiality, with the presence or absence of particular
properties being determined {\it in the process of} the evoking or
actualizing of the concept. In such situations, the mind handles
concept combination in a quantum manner.

\subsection{The Opinion Poll}
We extend the foregoing analysis of entangled states of concepts to a 
cognitive setting where an
individuals opinions are probed.  Opinions may appear to be more 
stable than other sorts of conceptual
entities that manifest as fleeting contents of conscious experience, 
such as impressions and ideas.
However, even opinions can exhibit  quantum-like contextuality. The 
formalism employed here for the
identification and description of nonclassical contextuality in 
cognition can be applied experimentally,
for instance, to the  analysis of opinion poles 
\cite{aertsaerts01,aertsaerts02}. To explain this we first have to 
introduce the
sphere-elastic model, which is a quantum model for the spin of a spin 
1/2 quantum entity \cite{aerts03,aerts06,aerts07,aerts08}.

The
states of the sphere-elastic model correspond to the points $P$ of 
the surface of a sphere, denoted $surf$, with center
$O$ and radius $1$. This means that we can denote a state $p_v$ by 
indicating the unit-vector $v$ corresponding to a point of $surf$ (see
Fig. 1,a), and the collection of all possible states is $\Sigma =
\{p_v\
\vert\ v \in surf\}$. For each point $u \in surf$, we introduce the 
context $e_u$ in the following way. We consider the diametrically
opposite point
$-u$, and install an elastic band of length 2, such that it is fixed 
with one of its end-points in $u$ and the other end-point in $-u$.
The influence of the context $e_u$ is the following. Once the elastic 
is installed, the particle $P$ falls from its original place $v$
orthogonally onto the elastic, and sticks to it (Fig. 1,b). The 
elastic then breaks and the particle $P$, attached to one of the two
pieces of the elastic (Fig. 1,c), moves to one of the two end-points 
$u$ or $-u$  (Fig. 1,d).

\vskip 0.5 cm

\hskip 0.5 cm \includegraphics{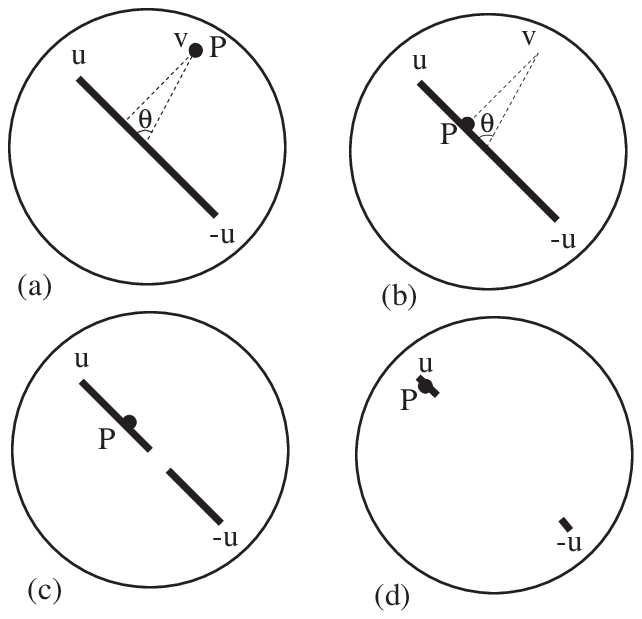}

\vglue - 5 true cm
\hangindent = 8 true cm \hangafter = - 9{\small \em \noindent  Fig. 1 
: A representation the sphere model. In (a) $P$ indicates a state
$p_v$ in the point
$v$, and the elastic corresponding to the context $e_{u}$ is installed between
the two diametrically opposed points $u$ and  $-u$. In (b) $P$ falls 
orthogonally onto the elastic and sticks to it. In (c) the
elastic breaks and $P$ is pulled  towards the point $u$,
such that (d) it arrives at the point  $u$.}

\vskip 2.2 cm

\noindent
The state $p_v$ is changed by the context
$e_u$ into one of the two states $p_u$ or $p_{-u}$. We make the 
hypothesis that the elastic band breaks uniformly, which means that
the probabilities of state transition under influence of the contexts are:
\begin{eqnarray}
\mu(p_u, p_v, e_u) &=& {{1+cos\theta}\over 2} = cos^2{\theta\over 2} \\
\mu(p_{-u}, p_v, e_u) &=& {{1-cos\theta}\over 2} = sin^2{\theta\over 2}
\end{eqnarray}
We can easily show that the sphere-elastic model is an entity of 
which the description is isomorphic to the quantum
description of the spin of a spin 1/2 particle. Hence, speaking in 
the quantum jargon, the sphere-elastic model is a
model for the spin of a spin 1/2 quantum particle. This means that we 
can describe it using
the ordinary quantum formalism with a two-dimensional complex Hilbert 
space as the carrier for the set of states
of the  entity. It is easy to see on the sphere-elastic model the 
effect of the lack of knowledge on the contexts: this corresponds to 
the
lack of knowledge of where the elastic will break during the 
interaction of the context with the state of the entity.

While in quantum mechanics the effect of context on the outcome is
complete, or coercive, in cognition it is intermediate, or tempered. 
The tempered quantum-like contextuality can
be described  using the $\epsilon$-model 
\cite{aertsetal08,aertsdurt01,aertsdurt02,aertsetal09}, which is an 
obvious generalization
of the sphere elastic model. For the $\epsilon$-model we introduce 
for each context $e_u$ a parameter $\epsilon \in [0, 1]$, and make the
hypothesis that the elastic corresponding to the context $e_u$ can 
now only break in an interval of length $2\epsilon$ around the middle
point of the elastic. This means that for $\epsilon = 1$, the 
$\epsilon$-model reduces to the original sphere-elastic model, hence 
a pure
quantum model, while for $\epsilon = 0$, each elastic can only break 
in its middle point, which means that we have a classical
situation, of contexts that have a deterministic effect on the state 
of the entity under consideration (with exception of what happens in
the middle point, but this is a classical type of indeterminism, as 
in the case of a classical unstable equilibrium position). For
intermediate values of $\epsilon$, smaller than 1 and greater than 0, 
the $\epsilon$-model describes the tempered situation of
quantum-like contextuality.

Let us describe shortly how the $\epsilon$-model can be used to model 
the quantum-like contextuality that appears in a opinion poll
situation. We consider three different questions for the opinion poll.

\indent  $u_1$ : ``Are you in favor of  the use of nuclear energy?"\\
\indent $u_2$ : ``Do you think it would be a good  idea to legalize 
soft-drugs?"\\
\indent $u_3$ : ``Do you think capitalism is better than  social-democracy?"\\
We have chosen typical questions on which many respondents  will not 
have predetermined opinions. Since the
respondent has to respond  with `yes' or `no', she or he, not having 
an opinion before the questioning
will `form' an opinion during the process of questioning itself. 
The $\epsilon$- model can be applied to
fingerprint this effect of creation in the opinion+context interaction.

\vskip 0.3 cm
\hskip 0 cm \
\includegraphics{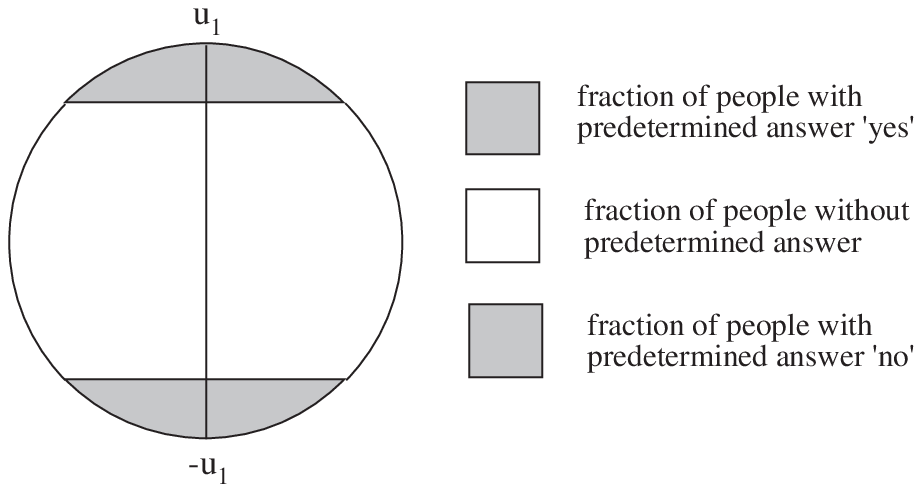}
  \vglue - 5 true cm
\hangindent = 12
true cm \hangafter = - 9{\small \em \noindent  Fig. 2: 
representation of the question $u_1$ by means of the 
$\epsilon$-model. We
have  indicated the three regions corresponding to predetermined 
answer `yes', without predetermined
answer, and predetermined answer `no'.}\\
\vskip  1.2  true cm

\noindent
To simplify the situation, but without touching the  essence, we make 
the following assumptions
about the probabilities that are involved.  We suppose that in all 
cases 50\% of the persons have answered
the question $u_1$ with `yes', but  only 15\% of the persons had a 
predetermined opinion. This means that
70\% of the persons formed their answer during the process of 
questioning.  For simplicity we make the same
assumptions for $u_2$ and $u_3$. We can represent this situation in 
the $\epsilon$-model as shown in Figure 2.
We also make some assumptions of the way in  which the different 
opinions related to the three questions
influence each other. One can see how a person can be a strong 
proponent for the use of nuclear
energy, while  having no predetermined opinion about the legalization 
of soft drugs (area 1 in Figure
3).  Area (4) corresponds to a sample of persons that have 
predetermined opinion in favor of legalization
of soft drugs and in favor of capitalism. For area (10) we have 
persons that  have predetermined opinion
against the legalization of soft drugs and  against capitalism. All 
the  13 areas of Figure 3 can be
described in such a simple way.

\vskip 0.3 cm
\hskip  0 cm \includegraphics{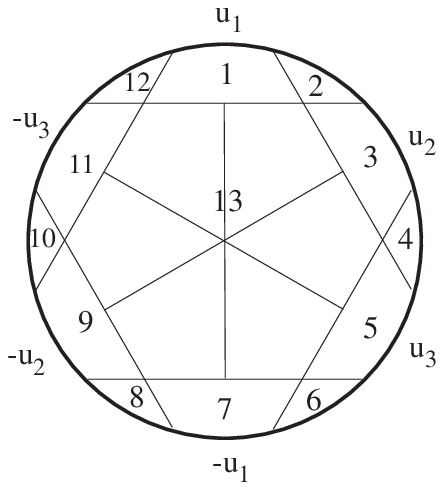}
  \vglue - 5 true cm
\hangindent = 5.5 true cm \hangafter = - 10{
   \small \em \noindent  Fig. 3: representation of the  three 
questions $u_1$, $u_2$, and $u_3$  by means of
the  $\epsilon$-model. We have numbered the 13 different regions. For 
example: (1)  corresponds to a
sample of persons that have predetermined opinion in favor of nuclear 
energy, but  have no predetermined
opinion for both other questions, (4) corresponds to a sample of 
persons that have predetermined opinion
in favor of legalization of soft drugs and  in favor of capitalism, 
(10) corresponds to a sample of
persons that have predetermined  opinion against the legalization of 
soft drugs and against capitalism,
(13)  corresponds to the sample of persons that have no predetermined 
opinion about non of the three
questions, etc...  }\\
\vskip 0.5  true cm
\noindent
Deliberately we have chosen the different fractions of people in such 
a way that the conditional
probabilities fit into the $\epsilon$-model for a  value of $\epsilon 
= {\sqrt2\over 2}$. It can be proven mathematically
that these values of conditional probabilities corresponding to these 
questions $u_1$,
$u_2$ and $u_3$ can neither be fitted into a Kolmogorovian 
probability model  nor into a pure  quantum
probability model \cite{aertsaerts01,aertsaerts02,aerts08}. This 
indicates that only part of the  properties are created during the
process of testing, and that the contextual influence  is therefore 
not fully coercive. If the respondents already had a fixed
opinion about  the question asked, the relevant probability model is 
Kolmogorovian and contextual
influence  is nihil.

This purely technical development of `interactive' statics does imply, with the
prerequisite of feasibility of  standardized experimental set up as 
described here, the classical analysis
of the opinion pole could be modified in order to identify 
opinion-context created effects.

\subsection{The Generalized Liar Paradox}
The abstracted quantum formalism is applied next to the modeling of a
compound  conceptual entity: {\em in casu} the generalized Liar
Paradox. When one reasons logically over a Liar Paradox --- e.g.
``this sentence is false"---, with an initial conceptualization of it
as `true', then one experiences a build-up of cognitive dissonance as
one recognizes by inference the inconsistency, which is followed by
its conceptualization as `false'. After which the logical oscillatory
process can be reiterated incessantly. We treat the conceptual entity
of the Liar Paradox  as a consistently testable configuration of truth
properties expressed by sentences, and subject to our capacity of
logical inference, which figures here as a coercive internal context.
These conceptual entities have the advantage of simplicity through
their autonomy relative to extraneous input other than the personal
dispositions of logical inference, e.g. contingent mind frame,
emotional state, environment.  And still the reflection on these
entities is non-trivial with respect to the dynamics of
self-reference.

The complete quantum formal description of the logical properties of
the Liar Paradox conceptual entity contrasts with the description of
the more generally encountered non-classical contextuality in the
previous subsections. Next to the non-deterministic contextual
collapse evolution, this model of the Liar Paradox allows the
introduction of the continuous deterministic evolution by reasoning
at any subsequent instance  of {\em time} as well. We detail the
coincidence of the effect of `inner' logical context and reasoning in
our concluding remarks.

Not only the simpler forms of the double
Liar Paradox \cite{aertsetal01,aertsetal02} but also
$m$-sentence generalizations \cite{aertsetal04} allow
a `complete' quantum description.
These particular entities---the
generalized $m$-sentence Liar Paradoxes---are e.g. with $m=5$:
\begin{eqnarray}
1 &   {\rm sentence\ 3\ is\ false }  &   \nonumber   \\
2 &   {\rm sentence\ 5\ is\ false }  &   \nonumber   \\
3 &   {\rm sentence\ 2\ is\ true  }  &   \label{5LP} \\
4 &   {\rm sentence\ 1\ is\ true  }  &   \nonumber   \\
5 &   {\rm sentence\ 4\ is\ false }  &   \nonumber
\end{eqnarray}
The first number is the sentence pointer, then follows the
corresponding proposition. The sentence-pointer numbers onto
proposition-content numbers,  and {\em vice versa},  link together
the sentences in a  closed `daisy chain' configuration.

In the present context we merely expose the basic components of the
formal model, while deleting most of the technical details 
\cite{aertsetal04}. The model represents the
state of the  Liar Paradox entity by tensor products of state vectors of the
sentence sub-spaces. Initially when the conceptual entity is in its
potential state no truth value of the entity should be explicited.
The truth and falsehood measurements on each sentence correspond to
appropriately chosen projectors, while each next step of the dynamics
is achieved by endorsing the
inferred truth value resulting in the ensuing projection.

We have shown the model needs to be constructed in a  $ ( 2 m )^m $
dimensional Hilbert space. However, the evolution of the given
paradox constrains the dynamics to a mere $2m$-dimensional subspace.
Given the choice of the truth and falsehood {\em by hypothesis}
operators, for sentence $i$:
\begin{eqnarray}  T_i &=&  {\bf 1}_1 \otimes ... {\bf 1}_{i-1}
\otimes T \otimes { \bf 1}_{i+1} ...\otimes {\bf 1}_{m}
\nonumber  \\ F_i &=&  {\bf 1}_1 \otimes ... {\bf 1}_{i-1} \otimes  F
\otimes { \bf 1}_{i+1} ...\otimes {\bf 1}_{m}
\label{vgloperators}
\end{eqnarray} with
\begin{eqnarray} T = 		{\scriptsize \left(
\begin{array}{cccc}
0   &  ... & 0   & 0    \\
  ... &  ... & ... &
...  \\
0   &  ... & 1   & 0    \\
0   &  ... & 0   & 0
\end{array} \right)}_{2m \times 2m}			 &{\rm and} &
F =
{\scriptsize \left( \begin{array}{cccc}
0   &  ... & 0   & 0    \\
   ... &  ... & ... & ...  \\
0   &  ... & 0   & 0    \\
0   &
... & 0   & 1    \end{array} \right)}_{2m \times 2m}  \label{TFop}
\end{eqnarray}
and  given a particular $m$-sentence configuration,
the corresponding reasoning sequence of $2m$ vectors can be
constructed.  All residual entries of the state vector are
unequivocally fixed by a `filling' procedure.   With respect to this
procedure, we obtained consistently interpretable  state functions in
terms of ``truth/falsehood by reference" and ``truth/falsehood by
hypothesis".

The first step of the reasoning on the Liar Paradox is represented
by letting a projector (\ref{TFop}), corresponding to the chosen
logical hypothesis, act on the initial superposition state $\Psi_0$,
an equiponderate expression of all possible truth and falsehood
states of component sentences. All subsequent steps of
inferences-hypothesis when reasoning through the generalized Liar
Paradox  are discrete in their effect and can therefore be conceived
as the operation of a step matrix.  In the  5-Liar Paradox example
(\ref{5LP}), this discrete
evolution submatrix is given by;
\begin{eqnarray}
{U_D}\vert_{\rm sub} &=&   {\scriptsize  \left( \begin{array}{cccccccccc}
0&{\bf 1}&0&0&0&0&0&0&0&0 \\
0&0&{\bf 1}&0&0&0&0&0&0&0 \\

0&0&0&{\bf 1}&0&0&0&0&0&0 \\
0&0&0&0&0&0&0&0&0&{\bf 1}
\\
0&0&0&0&0&{\bf 1}&0&0&0&0 \\
0&0&0&0&0&0&{\bf 1}&0&0&0 \\

0&0&0&0&0&0&0&{\bf 1}&0&0 \\
0&0&0&0&0&0&0&0&{\bf 1}&0 \\
{\bf
1}&0&0&0&0&0&0&0&0&0 \\
0&0&0&0&{\bf 1}&0&0&0&0&0 \\
    \end{array} \right) }
\end{eqnarray}
The discreteness in the temporal process features explicitly in the
logical reasoning. In the present model, the formalism of quantum
mechanics on the other hand allows a continuous time parameter of
evolution, allowing `intermediate'
states of   reasoning on the Liar Paradox. This is done by extracting
the phenomenological Hamiltonian---the infinitesimal time
propagator of the system---through the application of Stone's
Theorem (see Figure 4). 

In the full
quantum model an initial  hypothesis about one sentence  engenders a
time evolution of build up and collapse of
logical states without end.   Evidently any real world reasoning on
the Liar paradox does not expose this compulsory
machine-like continuation of the process.

Then what does the complete quantum description reflect? The autonomy
of dynamics does not necessarily intend an ontological reality of the
isolated conceptual entity. The latter could be inferred from too
literal interpretation of the physical
analogy with the obtained complete quantum description. On the other
hand, the  construction  procedure of the step
evolution reflects the cognitive person's motivation by reasoning.
The obtained dynamical model extends the  `entity +
context' configuration which is static. 

We have therefore been able in
this case to introduce a time-propagator
characterizing the concatenation of states of thought, albeit
restricted to a non-evolving  coercive logical context of
logical inference. It is to be expected that more general conceptual
entity with variable internal context and environmental context will
not fit a complete quantum model, {\em a fortiori} one with a
permanent propagator. We can interpret the descriptive coincidence of
the `autonomous evolution of an ontological
state' and  `coercion by inner logical context' as the mechanism that
lends the conceptual entity its {\em intentionality}.
The conceptual entity `liar paradox' refers over time and
consistently to the inner context of logical inference, which
is indeed, in this particular case, sufficient ground for recognizing
its intentionality.

\vskip  - 0.8  true cm
\hskip - 0.5  true cm
\includegraphics{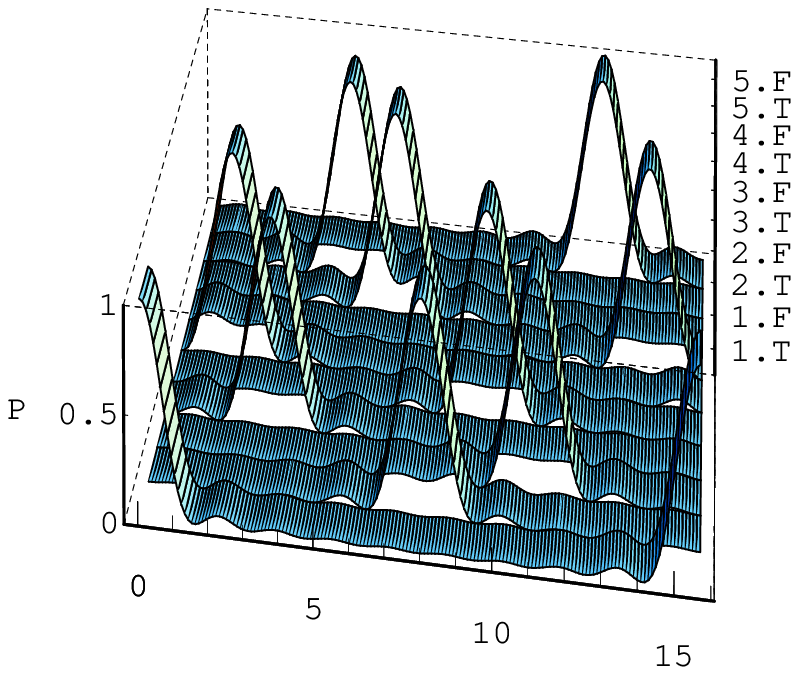}
  \vglue - 7  true cm
\hangindent = 8.5  true cm \hangafter = - 19 {
   \small \em \noindent Fig. 4: {\bf Time evolution} of outcome probabilities
for reasoning
the 5-Liar Paradox (\ref{5LP}), with at $t = 0$, a `true'-measurement
of sentence 1  on the initial state $\Psi_0$. The `time'  $t$ is an
arbitrary continuous ordering parameter without physical
interpretation. Logical contradiction  is apparent after each
interval $\Delta t = 5 \frac{\pi}{2}$. The probability  for a given
outcome ``sentence $i$ is $T/F$  at time $t$" is obtained  by taking
the modulus squared of the sca\-lar product  of sentence of substate
``$i$ is $T/F$" and the  initial state evolved till time $t$, i.e:
$P(i.{\scriptsize T/F}, 1.T, t )=  \left\vert \left< \psi_{i.{\scriptsize T/F}}
\vert U (t) P_{1.T} \Psi_0 \right> \right\vert^2$. }
\vskip  0.7  true cm

\noindent
Finally, does the quantum model `solve' the liar paradox? The feature
of the non-classical
contextuality reflects a distinct ontological status of the Liar
Paradox as a highly contextual entity when  compared to
conceptual entities with fixed classical truth or falsehood. The
model therefore suggests that entities exist which are not subject to
classical logical categories. When we try to understand the paradox
through these classical categories, we are caught in a contradiction
which, in this model, can be avoided by separation of incompatible
values over time.

\subsection{Thought as a Cognitive Form of Context-driven
Actualization of Potential}
As we mentioned in section \ref{sec:classicalquantumcontext}, in
quantum mechanics there are two fundamental forms of change: (1)
continuous dynamical evolution in the absence of a measurement as
described by the Schr\"odinger equation, and (2) quantum collapse
when a measurement takes place. We propose that conscious experience
similarly consists of two phases: (1) a continuous, predictable
unfolding of states when the situation does not require a decision to
be made and presents no outstanding ambiguity, and (2)
context-driven collapse of the conceptual network when a decision
must be made, or an ambiguous stimulus or situation resolved
\cite{gabora01}. In this second situation, we say that prior to the
decision or resolution of ambiguity, the mind is in a state of
potentiality. This is demonstrated schematically in Figure 5, which
shows the potentiality state of the concept `diamond' and the various
different contexts that actualize to differing degrees the various
different properties of `diamond'. Thus, for example, in the context
of wanting to cut metal, the property `strength' would be actualized,
whereas in the context of wanting to marry someone, the property of
being a symbol of love would be actualized. In the context of wanting
to make money, the property `valuable' is actualized. This context
might pull in properties that are not generally associated with
diamond, such as that a fake diamond can be constructed from plastic.

\vskip 0.4 cm
\hskip 0 cm \
\includegraphics{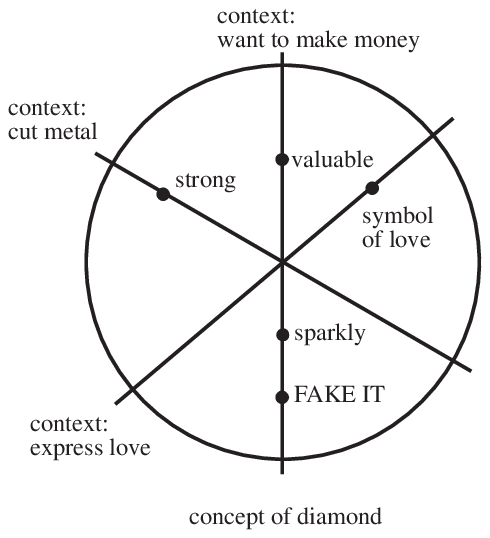}
  \vglue - 5 true cm
\hangindent = 8 true cm
  \hangafter = - 12{\noindent  \em Figure 5. The potentiality state 
associated with the concept `diamond' showing some of the many 
features that could potentially be present in various different 
context-driven actualizations of this concept.
.
   \small \em \noindent  }\\
\vskip  4  true cm
\noindent
We can view situations as `testing' or measuring the state of the
conceptual network, revealing something of its hidden structure by
showing how it collapses in response to a certain kind of context.

\vskip 0.7 cm
\hskip 0 cm \
\includegraphics{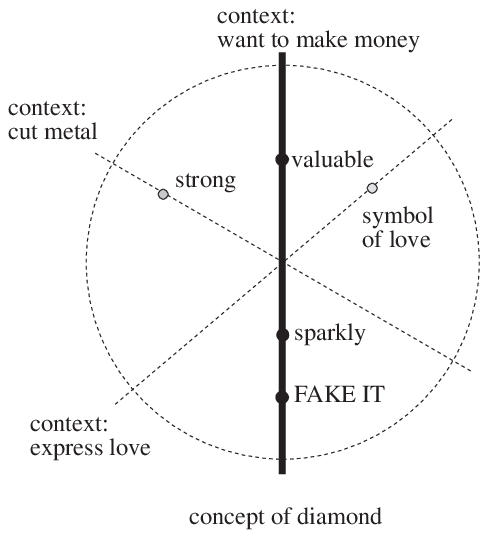}
  \vglue - 5 true cm
\hangindent = 8 true cm
  \hangafter = - 12{\noindent  \em Figure 6. The collapsed 
state of the concept `diamond' in the context
of wanting to make money. Some of the features that were previously 
potential are no longer potential, whereas others are now actualized.
   \small \em \noindent  }\\
\vskip  2.8  true cm
\noindent
The collapse is demonstrated schematically in Figure 6. In this case,
the context is that of wanting to make money. This context causes the
concept `diamond' to collapse from a superposition type state where
many elements are {\it potentially} relevant to an `eigenstate' (end
state) where specific elements are {\it actually} relevant.

More broadly, both the continuous and discontinuous aspects of both
conscious experience and quantum
mechanics can be viewed as different means of instantiating Context
Driven Actualization of Potential or $CAP$ \cite{gabora01}. Both of
the continuous forms of change can be seen as the degenerate case of
the discontinuous; with respect to the present context there is
only one possible outcome, thus the outcome is determined. Since the
potentiality is limited to one outcome there is no need for
collapse; this one alternative just plays itself out.

It is interesting to note that two of the greatest stumbling blocks
we now face in the formalization of conscious processes are extremely
similar to the two main problems that have been identified in quantum 
mechanics. These
problems  can be referred to as the {\it measurement problem} and the {\it
entanglement problem}.

\begin{itemize}
\item {\it The measurement problem}:
Classical mechanics cannot describe situations where the measurement
affects the state of an entity. For quantum mechanics the effect of 
change provoked by the measurement
context on the state of the quantum entity is an essential ingredient 
of the theory. Similarly, when a
concept (e.g. `diamond') is evoked, it is evoked in some context, 
which affects its  state. Therefore
we need a formalism that incorporates context into  description of entity.

\item {\it The entanglement problem}:
If in classical mechanics the composed entity consisting of two 
entities is described, this
composed entity does not contain`new' states, that cannot be reduced 
to product states. In quantum
mechanics this is essentially the case, the composed entity 
consisting of two quantum entities,
always contains `new' states, the so called entangled states. Similarly,
in creative  process, concepts merge to generate something new with 
properties not reducible to the
given constituents. Again, we need a formalism for  describing how 
two entities combine into one and
how new states are formed during this combination.

\end{itemize}
The similarity of the problems faced in the two domains should make
it clear that the application of formalisms from one to the other was
not pulled out of a hat. In fact, it is
part of an effort to gather different forms of change---physical,
chemical, geological, biological, cognitive, cultural, social,
economic, and so forth---under one umbrella that identifies and
formalizes their similarities and differences with respect to: (1)
Degree of contextuality, (2) Degree of indeterminism due to context,
(3) Degree of context dependence, and (4) Degree to which
context-driven change is retained in future lineage(s). For present
purposes it would not be appropriate to go into this general
framework in detail. Our point is merely to show that the application
of quantum formalisms to cognition is part of an
interdisciplinary effort to develop a general framework for comparing
and contrasting processes of change.

\section{The Ontology of Cognition and Consciousness}
We now outline the relationship of the developments of abstracted
quantum formalism in cognition described here to the general
question of the ontology of cognition and consciousness. This
approach provides possibly a sufficient framework for a process of
consciousness in systems with the capacity of
conceptualization. We will juxtapose this with an {\em
amplification} process model of consciousness \cite{gabora02}, which is
also consistent with the presented intrinsic contextuality formalism, but
extends it with the additional assumption the consciousness is a 
fundamental primitive.

\subsection{The `Reflected' Internal Context Model of Consciousness}

The abstracted quantum formalism we apply to cognition sustains a
view of consciousness as an emergent, integrated process involving 
reflection of the entity via the
context. We now reiterate the underlying arguments for and expand 
upon this view.

First we mention that the presence of quantum-like probability structure
is not a sufficient indication for emergent processes in the
conceptual entity-context  interaction, {\em i.e.} with genuine new
states resulting. The quantitative tool of statistics discerns
classical and quantum-like probability structure, given the
constraints of  multiple states and multiple contexts, and does so
depending on contextual uncertainty or interaction uncertainty.  This
leaves open a number of ontologically very different decisive factors
for the quantum-like probability structure: `fuzzy' context,
fluctuating interaction, and indeed entity-integration.
An interaction process giving rise to a nonKolmogorovian probability
structure  can still have an underlying deterministic mechanics.
Potentially this type of dynamics provides emergence but fails to
procure the new state properties in terms of the created entity, and
shunts the typical features of the quantum approach in which it is
embedded (cf below). On the other hand an emergent integration
process of conceptual entity+context will provide the necessary
supplementary variability---it has new features---in the interaction
to certainly break the Kolmogorovian axioms, and thus lead to the
quantum-like probability structures. But this effect is not
unequivocally expressed since the quantum-like structure could
already have been accounted for by mere contextual uncertainty. So we
find in quantum-like  probability structure no decisive indication
of emergence.

More potent arguments come from the various examples of
quantum-like features in cognitive processes using different quantum
approaches (Bell, Schr\"odinger, $\epsilon$-model, CAP evolution);
these do strongly suggest an emergent process. In
quantum physics the non-deterministic change , {\em i.e.} the
measurement collapse, has an extra-ordinary function: it relates the
single individual entity to the coercive superstructure of the
measuring context. The measurement collapse thus integrates hybrid
components, formally it subjects the individual entities state
function to the projection operator of the context. Physically
although we understand that a connection between both is possible;
the superstructure of the context is in fact a highly structured and
organized compound system of elementary entities. The architecture
and organization of the superstructure supplies it with supplementary
functionality, the acquired `new' properties. These emergent features
can be understood in the manner of complex dynamics (a  process
widely accounted for in the literature, e.g. \citeasnoun{scott01}). The von
Neumann measurement collapse is therefore a generic procedure and
formalism for the interaction of  an entity and its relative
superstructure. In cognition exactly the same relation is present
between the conceptual entity and the cognitive context. The
effective context in the cognitive scheme refers to a vast network of
related concepts, memories, drives, stimuli and supplemented by the
induced effects of the external environment, which constitute finally
the present mind frame. The cognitive context shows to some extent
the same hybrid relation with  conceptual entities and has to some
extent the same hybrid coercive effect. Only on these grounds we can
identify the quantum-like features in our model with emergent
processing.

Finally we try to relate parsimoniously the `intrinsic
contextuality approach to cognition' to a model of
consciousness.  From a conservative point of view, consciousness
indicates a capacity `to perceive the being in the world',
as such reducing it to a cognitive state be it a centric one. To
analyze  this process in terms of the intrinsic contextuality
approach. As a prerequisite, we let conceptualization be an acquired
functionality of a context stimulated, feedback regulated,
consistently referent, activity of the neural network.  Attention is 
directed by a number of factors which
are expressed in the  internal context and the shift of this context, 
and provides the
present cognitive experience.  Their relatively hybrid nature
integrates and  procures the cognitive experience. To extend
cognition to consciousness is then to direct the mental context  to
this proper context itself as subject.  In common cognitive tasks, the
mental context is in an emergent relation to the conceptual entities,
the refocusing in consciousness does not fit that common
configuration. The practical `impossibility' to subject the mental
context to itself in an act of thought, indicates to some extent the
evanescent nature of consciousness. In effect such an attempt
probably  converts to experiencing these components of the internal
context that relate to proper existence.

The present elementary
and speculative account for the process of consciousness is based
directly on the intrinsic contextuality model in cognition and its
translation of  `proper reflection'.  Its merit lies in the
consistency with the ontology of precedent concept+context analysis
and its parsimonious plainness. This  account puts the essential
dynamics of cognition and consciousness in the context directing
capacity of the mind,  a process not accounted for here.

\subsection{The Amplification Model of Consciousness}
Fundamental approaches bypass the problem of getting
consciousness from non-conscious components by
positing that consciousness is a universal
primitive
\cite{chalmers01,dyson01,feigl01,edelmantonini,forster01,ghoseaurobindo01,griffin01,hartshorne01,lockwood01,montero01,nagel01,russel01,scott01,stoljar01,strawson01,whitehead01}.
The double aspect theory of information,
for example, holds that information has a  phenomenal aspect 
\cite{chalmers01}. How then do you get from
phenomenal information to human consciousness? Our emphasis here on 
the interface between cognitive state
and context might appear to suggest  we believe that consciousness is 
uniquely associated with a
particular kind of cognitive, or perhaps organic, structure. Our
position, however, is consistent with the view that subjectivity of
some very primitive form is ubiquitous, and that these structures
merely locally amplify this subjectivity to give rise to what we
generally view as full-fledged consciousness \cite{gabora05,gabora02}. It has
been proposes that an entity is conscious to the extent it amplifies
information, first by trapping and integrating it through closure,
and second by maintaining dynamics at the edge of chaos through
simultaneous processes of divergence and convergence. The origin of
life through autocatalytic closure induced phase transitions in the
degree to which information, and thus consciousness, is locally
amplified. Another such phase transition may have taken place with
the origin of an interconnected worldview through conceptual closure
\cite{gabora03,gabora04,gabora02,gaboradeses01}.

Conceptual closure is a process whereby memories become
increasingly able to evoke one another mediated by the formation of
abstract concepts, and eventually form an interconnected internal
model of the world. It has been proposed that the
capacity for conceptual closure came about approximately two million
years ago when the human memory became large enough that episodes
could afford to be more widely distributed, leading to the formation
of a hierarchical network of abstract concepts. It has also been
proposed that the capacity to alternate between focused and defocused
modes of thought led to the capacity for conceptual closure at
multiple hierarchical levels spanning different domains, and that
this is what gave rise to the creative revolution of the Upper
Paleolithic. Such a conceptual structure is able to creatively
fine-tune plans, predictions, ideas and fantasies by evaluating
situations in terms of past experiences, stories, and schemas, update
opinions by considering them from different perspectives, and merge
concepts to generate new concepts (which, liked quantum entities,
often have properties not present in the constituents). All of these
involve contextuality, revising one thing by viewing it in the
context of another. Thus we move toward a picture in which closure,
quantum structure, and contextuality are three intimately connected
aspects of the process through which consciousness has become
amplified.

At the deep level of formal theories, there is in fact a tight
connection between quantum structure, and closure, which we discuss
here briefly to give the reader a flavor of what this could imply for
cognition. We frame this explanation using the lattice theoretic
generalized quantum formalism, because it is here that this
connection is revealed most clearly, though since the generalized
quantum formalisms are translatable one into the other, the
connection between closure and quantum structure could be analyzed
using any one of them. In the lattice approach, the basic
mathematical concepts for describing a physical entity, whether it be
purely classical, purely quantum, or of the more general sort of
structure referred to as quantum-like, are (1) a state space and (2)
a set of properties. Axioms are defined on the state space and the
set of properties, thus turning the whole into a state-property
system \cite{aerts04,aerts05}. The entity under consideration is a 
classical entity if
the lattice is Boolean, and pure quantum if the lattice is
irreducible. It can be shown that if all the axioms are satisfied,
the state-property system can be represented as the direct union
over a set $\Omega$ of lattices formed out of the closed subspaces of a
generalized Hilbert space \cite{aerts02,aertsetal05,aertsdeses01}.

The pure classical situation corresponds to the case where all these
generalized Hilbert spaces have dimension one, hence this direct
union reduces to the set $\Omega$ itself, which represents the phase
space of the classical entity. So the general theory reduces to a
pure classical mechanical theory over phase space. The pure quantum
situation corresponds to the case where the set $\Omega$ is a singleton,
hence the direct union reduces to one generalized Hilbert space,
which for the case wherein the division ring is the complex number
field, reduces to the Hilbert space of pure quantum mechanics. So the
general theory reduces to a pure quantum mechanical theory over
Hilbert space. The general quantum entity is described by a direct
union over a phase space $\Omega$ of closed subspaces of generalized
Hilbert spaces. It is $\Omega$ that enables one to describe the
classical aspects of the entity, and the closed subspaces of
generalized Hilbert spaces that allow for description of the quantum
aspects.

It was recently proven that a state property system in the general
quantum situation is categorically isomorphic with a closure
structure on the state space 
\cite{aertsetal06,vandervoorde01,vansteirteghem01,aertsetal07}. The
connected components of this closure  structure are the pure quantum components 
of the generalized
Hilbert  spaces in the direct union, while the disconnected components are
distributions over the phase space $\Omega$ that represents the
classical aspects of the considered entity 
\cite{aertsetal05,aertsdeses01}. Thus it appears that, at
a deep formal level, the connected components of a closure system are
the quantum components of that system, and the disconnected
components are the classical aspects. With respect to cognition, this
tentatively suggests that closure of the conceptual network which
causes it to become interconnected by hierarchical levels of
increasingly abstract concepts, causes the mind to acquire more
quantum-like structure.

\section{Conclusions}
The contextual nature of conscious experience suggests that in order
to formally model it we should look to the domain of science where
contextuality has been most seriously addressed: quantum mechanics. 
We know  that conceptual entities {\it always} get evoked in some 
context. However the
incorporation of context into the formal description of a conceptual 
entity is  one of the biggest unsolved problems in cognitive science. 
Our
formalism lends itself precisely to this problem.

An abstracted quantum mechanical representation of the entity-context 
interaction, with its hidden creation of new states, was adapted to 
the
description of the conceptualization process for various different 
kinds of cognitive situations. We showed that Bell inequalities---the 
definitive
test for quantum structure---are violated in the relationship between 
an  abstract concept and instances or exemplars of that concept.

  A formal
$\epsilon$-model derived from quantum mechanics was adapted to the 
setting of an opinion pole, and identifies possible effects of 
intrinsic
integration of opinion and questioning context, {\em i.e. 
contextually-elicited opinions}.

The full quantum model of the specific conceptual
entity the `liar paradox'  provided a case for complete coercive 
contextuality and time evolution of a contextually subjected entity.

As in
quantum mechanics, conscious experience consists of segments of 
dynamical evolution, which are not contextual and do not involve 
resolution of
ambiguity or decision, and collapse events, which are 
context-dependent and involve a decision or the resolution of 
ambiguity. In both the quantum
and the cognitive situation, both dynamical evolution and collapse 
are instances of context-driven actualization of potential. We 
believe that this
actualizing of potential through contact with a context is the 
fundamental basis of evolution and change.

Our application of the general formalism still needs to be worked out 
in greater detail, and the predictions of the resulting
`contextualized' theory  of cognition need to be empirically tested 
against other theories (such as prototype and exemplar theories of 
concepts),
both through  experiments with human subjects, and through 
simulations. We are also working on a description of the process by 
which
pre-existing concepts or ideas merge to form new ones using the 
mathematics of entanglement and collapse.

In conclusion, we believe consciousness to be an emergent, integrated 
process involving reflection of the entity via the context, which can 
be
fully described using an abstracted quantum formalism, and that 
amplification through closure may also be involved.

\end{document}